\documentclass[prb,notitlepage,twocolumn,authoryear,superscriptaddress]{revtex4-1}
\usepackage{epstopdf}

\usepackage{amsmath}
\usepackage{widetable}

\usepackage[pdftex]{graphicx}
\usepackage[usenames,dvipsnames]{color}
\usepackage{amssymb,braket}
\usepackage{ulem}

\usepackage[colorlinks=true,linkcolor=black, citecolor=blue, urlcolor=blue]{hyperref}
\usepackage{mathtools}

\usepackage[colorinlistoftodos,prependcaption,textwidth=1cm,textsize=footnotesize,color=lime]{todonotes}

\newcommand{\Imag}{\mathcal{I}m{}}%imaginary part
\newcommand{\1}{$_1$}%for F\1 = F$_1$.
\newcommand{\2}{$_2$}%for F\2
\newcommand{\3}{$_3$}%for F\3
\newcommand{\4}{$_4$}%for F\4
\newcommand{\5}{$_5$}%for F\5
\newcommand{\Ic}{j_c}%critical current density

\definecolor{green}{rgb}{0,.5,0}

\newcommand{\pt}[1]{\textcolor{Plum}{{\bf #1}}}

\begin{document}

\title{The role of canting and depleted-triplet minima\\ in superconducting spin valve structures}
\author{Thomas E. Baker}
\affiliation{Institut quantique \& D\'epartement de physique, Universit\'e de Sherbrooke, Qu\'ebec J1K 2R1 Canada}
\affiliation{Department of Physics \& Astronomy, University of California, Irvine, California 92697 USA}
\author{Andreas Bill}
\email[Author to whom correspondence should be addressed: ]{andreas.bill@csulb.edu}
\affiliation{Department of Physics \& Astronomy, California State University Long Beach, California 90840 USA }
\date{\today}

\begin{abstract}
The trilayer and pentalayer spin valve structures are revisited to determine the behavior of pair correlations and Josephson current when the magnetic layers are canted at arbitrary angle.  The two systems display markedly different behaviors in the center magnetic layer. While the trilayer generates a triplet component that is weakly affected by canting, the pentalayer tunes in singlet pair correlations depending heavily on canting. We also show that a minimum with depleted $m=\pm1$ triplet components, rather than a $0-\pi$ transition, may be observed in the current profile $I_c(d_F)$ of a trilayer spin valve.  The depleted-triplet minimum (DTM) is directly attributable to a decrease of $m=\pm1$ triplet correlations with increased thickness of the central ferromagnet, accompanied by a  hidden, simultaneous sign change of the Gor'kov functions contributed from the left and right superconductors.
 We introduce a toy model for superconducting-magnetic proximity systems to better illuminate the behavior of individual components of the Gor'kov function and compare with a full numerical calculation.
 
\end{abstract}

\maketitle

\section{Introduction}\label{s:intro}
\begin{figure}%[rb]
\begin{center}
\includegraphics[width=.8\columnwidth]{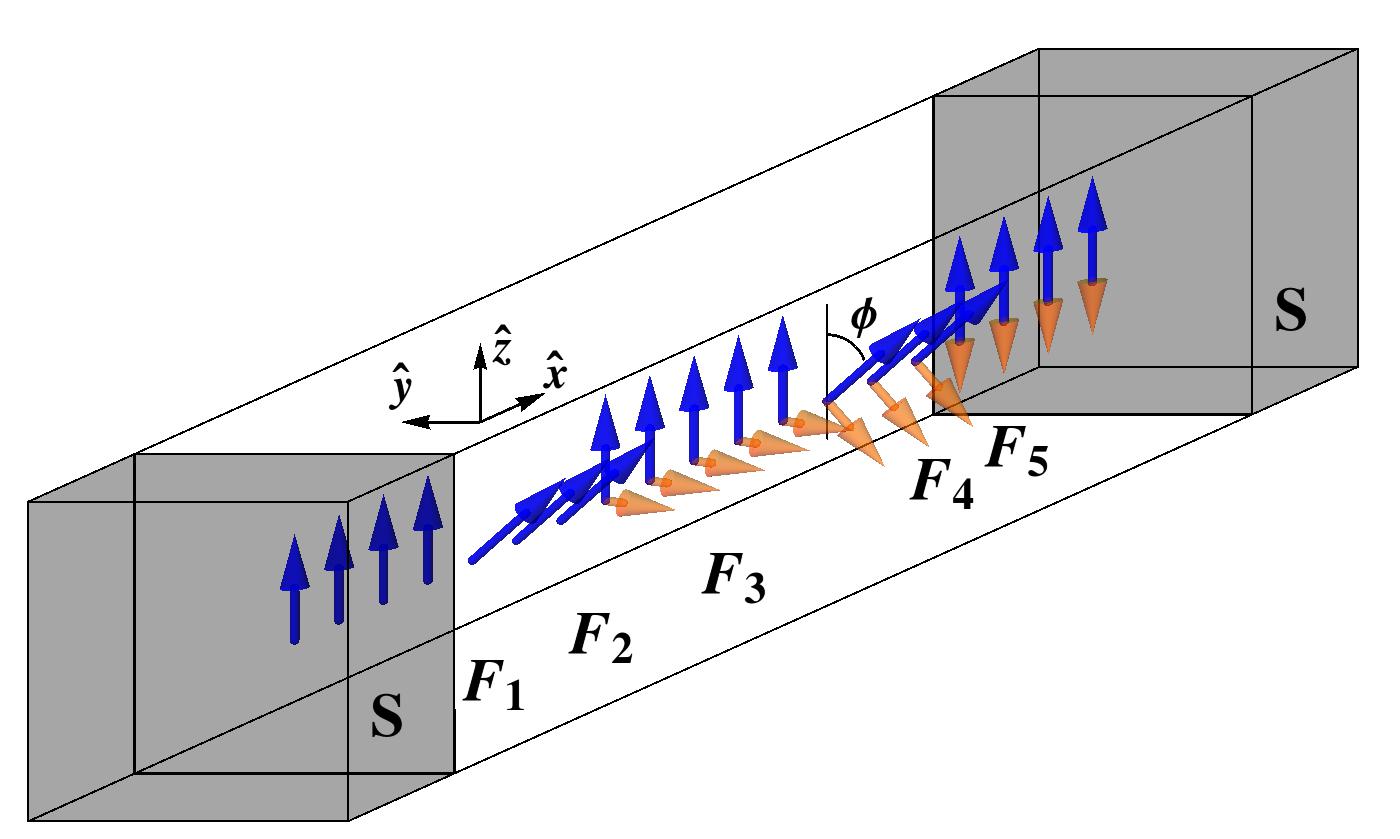}
\end{center}
\caption{(color online) \label{cantedfigure} Shown in dark blue is the magnetization profile of a canted 5F spin valve system with F\2 and F$_4$ canted by an angle, $\phi$. \cite{Note1}  Translucent orange arrows show the canted Bloch-like domain wall  magnetization ($\pi$-flip or $\pi$-wall).  We also consider a spin valve with three F layers which is realized by aligning F$_2$, F$_3$ and F$_4$.}
\end{figure}

A spin valve consisting of more than one homogeneous ferromagnet (F) in proximity with a singlet pairing superconductor (S) has been a popular tool with both experimentalists and theorists to explore odd frequency triplet pair correlations in magnetic hybrid systems.\cite{bergeretPRB01,houzetPRB07,khairePRL10,anwarPRB04,robinsonS10,eschrigRPP15} Creating a heterostructure with tunable properties is highly desirable in spintronic applications.\cite{linderNP15,gingrichNP16} and understanding pair correlations in these hybrid structures is of current experimental interest.\cite{martinez2015amplitude,eschrigRPP15} Before being destroyed by the exchange field of the F, singlet Cooper pairs from S may acquire an angular momentum, $s$, generating triplet correlations with $m=0$ (in the $\ket{s,m}$ basis of spin-$1/2$ fermions pairs) through the Fulde-Ferrel-Larkin-Ovchinnikov (FFLO) effect.\cite{fuldePRL64,larkinJETP65}  Rotating the magnetization's direction in space changes the natural quantization axis and causes mixing of zero-spin-projection $m=0$ and parallel-spin $m=\pm1$ pair correlations.\cite{bergeretPRL01,kadigrobovEPL01}
The main feature of the trilayer (3F) spin valve  (see Fig.~\ref{cantedfigure} with F$_2$,F$_3$,F$_4$ aligned) is that with a judicious choice of layer thicknesses, the $m=0$ components decay in amplitude to negligible values in the middle layer so that measured quantities are predominantly determined by $m\neq0$ components.\cite{keizerN06,khairePRL10,anwarPRB04,robinsonS10,klosePRL12,zhuPRL10,khasawnehSST11,wenEPL14,leksinPRL12,gingrichNP16,glick2017spin,martinez2015amplitude}
Recently, a calculation made for a pentalayer (5F) spin valve with $\pi/2$ orientation of the magnetization in F$_2$, F$_4$ showed that $m=0$ components are recovered far beyond one coherence length of an SF or FS interface.\cite{bakerEPL14}  An experimental test of the presence of these unexpected $m=0$ correlations was proposed.

In this paper, we investigate the 3F and 5F geometries for various canting angles (see Fig.~\ref{cantedfigure}),\footnote{Note that the angle $\phi$ is {the complementary of the angle $\phi'$ in Ref.~\onlinecite{bakerNJP14,bakerEPL14} ($\phi=\pi-\phi'$).}} and explore the effects of canting on the Josephson current. The analysis shows that while canting does not very much affect the pair correlations mixture in the 3F geometry, these correlations are strongly canting dependent in the pentalayer 5F. Surprisingly, the most drastic changes occur at small deviations from the (anti) parallel or perpendicular configuration of neighboring magnetic layers ($\phi$ slightly off $0,\pi/2$, or $\pi$) and is steadier farther away from these values. 

Further, we demonstrate the presence of a new, direct signature of $m=\pm 1$ pair correlations in the Josephson current that may be of interest for applications. A dip, which is a depleted-triplet minimum (DTM), of the current is observed as a function of the central F layer thickness. This DTM of the current is related to two features of pair correlations: a) The minimum of the current is found when the average position of all Matsubara frequency nodes in the $m=0$ Gor'kov functions coincides with the interface between two Fs; b) The Gor'kov function generated from the left and right Ss {\it simultaneously} change sign at the thickness of the minimum of the dip. This latter feature contrasts the DTM from the $0-\pi$ transition of the Josephson current since the latter is seen when only one of the two Gor'kov functions changes sign at the position of the dip, thereby turning into a node.
Here we show this effect in a trilayer spin valve as one varies the thickness of the outermost F layers.  The DTM is a general feature of hybrid systems that is only related to the two properties stated above and thus should be visible at any canting and in any multilayer system, as long as the thicknesses are chosen appropriately.

Section~\ref{s:method} discusses the methods used for numerical calculations presented throughout the paper.  Section~\ref{s:trilayer} discusses the trilayer, starting with results for the pair correlation functions on a range of canting angles in Sec.~\ref{s:trilayerPairs}.  The relation to the current is discussed in Sec.~\ref{s:trilayerJc}.  The same discussion is conducted for the pentalayer in Sec.~\ref{s:pentalayer}, focusing on the pair correlation functions in Sec.~\ref{ss:pentalayerPairs} and Josephson current in Sec.~\ref{ss:pentalayerJc}. The DTM is discussed in detail in Sec.~\ref{s:toymodel} by introducing a toy model to fully understand the effects.  Sections~\ref{s:trilayerPairstoy} and \ref{ss:trilayerJctoy} discuss the pair correlation functions and Josephson current, respectively.

\section{Methods}\label{s:method}
 
We conduct our analysis in the diffusive regime, where the elastic scattering length is much less than the coherence lengths in the system, and Usadel's equations apply.\cite{usadelPRL70} These equations and the approach to solve them numerically at finite temperature have been described in Refs.~\onlinecite{bakerNJP14,bakerPRB16}.
We only point out here that we use the Matsubara formalism, and
the Green functions (expanded as $\mathcal{G}=g_0+\mathbf{\hat v}\cdot\mathbf{g}$ with Fermi-velocity $\mathbf{\hat v}$) and Gor'kov functions $(\mathcal{F}=f_0+\mathbf{\hat v}\cdot\mathbf{f})$
are parameterized by trigonometric functions following Refs.~\onlinecite{ivanovPRB06,ivanovPRB09} (see also Refs.~\onlinecite{bakerNJP14,bakerPRB16} for implementation details). In F the superconducting pair potential, $\Delta$, is zero while $\mathbf{h}(x)$ is the position dependent magnetization profile with magnitude $h$. The latter is zero in S.

We introduce the coherence length $\xi_c=\sqrt{D_F/(2\pi T_c)}$, with critical temperature $T_c$ of the proximity system and diffusion length $D_F$ of the F, in order to compare different Fs on the same length scale. Other length scales are $\xi_F=\sqrt{D_F/h}$ that characterizes the decay of $m=0$ components, and the normal state coherence length $\xi_N=\xi_c\sqrt{D_N/T}$ (at temperature $T$) over which singlet pair correlations decay in a normal metal and $m=\pm1$ components decay in a F.\cite{buzdinRMP05,billJSNM12} Note that $h\gg T$ typically and thus $\xi_F$ is only a few nanometers even in a weak F whereas the $m\neq 0$ components may propagate at length scales of the order of $\xi_N$ that are much larger.

We consider transparent interface conditions between Fs, where the values and derivatives of the functions match on either side of each interface, noting
though that the transparency can affect the results.\cite{vasenkoPRB08}  At the SF interfaces, the boundary condition is set to $(M_0,\mathbf{M})=(1,\mathbf{0})$.\cite{parametrization}  The boundary condition on $\vartheta$ (the trigonometric functions of  Ref.~\onlinecite{bakerPRB16}; see Ref.~\onlinecite{parametrization}) is the bulk value in S, $\vartheta(\mathrm{SF})=\vartheta(\mathrm{FS})=\theta_B=\arctan(|\Delta|/\omega_n)$ ($\omega_n$ is the fermionic Matsubara frequency; see Ref.~\onlinecite{bakerPRB16}).

In the wide limit, the Gor'kov function $f(x,\omega_n) \equiv f_n(x)$ (for each $\alpha = 0,y,z$ we have $f_{\alpha,n}(x)$) may be written as the sum of two components from the left SF proximity system ($L$)
and right FS system ($R$)
superconductors\cite{vasenkoPRB08,belzigS&M99}
\begin{equation}\label{eq: gorkovadd}
f_n(x)=e^{i\varphi/2}f_{n,L}(x)+e^{-i\varphi/2}f_{n,R}(x),
\end{equation}
where $\varphi$ is the phase difference between the Ss. Hence, the contributions from the left and the right S may be calculated separately and added together.  Calculating the components generated by each S (L and R) independently also allows for a clear representation of pair correlations in the proximity system.

Once the Gor'kov functions have been obtained, the measurable Josephson current density is\cite{houzetPRB07,champelPRL08}
\begin{equation}\label{jcdensity}
\Ic(x)=\frac{\pi T}{2eR_N}\sum_{\omega_n\geq 0}\sum_{\alpha=0,y,z}\Imag[f^*_{\alpha,-n}\partial_xf_{\alpha,n}],
\end{equation}
requiring an integration over the thickness of F for the total current $I_c$ ($e$ is the electron charge, $R_N$ is the normal state resistance).\footnote{A position dependent spinor that has significance only near the SF or FS interfaces has been neglected.
Hence, the analysis best applies to wide F as in Refs.~\onlinecite{zaikinZP81,belzigS&M99,vasenkoPRB08}}

\section{Trilayer Spin Valves and canting}\label{s:trilayer}

\subsection{Effect of canting on pair correlations}\label{s:trilayerPairs}
We consider SF$_1$F$_2$F$_3$S with the following magnetization profile
\begin{equation}\label{3Fmag}
\mathbf{h}(x) = 
\begin{cases}
h\, \mathbf{\hat z},&x\in F_1, F_3,\\
-h \sin\phi \, \mathbf{\hat y} + h\cos\phi \, \hat{\mathbf{z}} ,&x\in F_2,
\end{cases}
\end{equation}
where $\phi$ denotes the arbitrary but fixed, constant angle of the magnetization in F$_2$ with respect to $\hat{\mathbf{z}}$.  The origin of the coordinate system has been set at the center of the magnetic multilayer (see Fig.~\ref{cantedfigure}).
\begin{figure}
\includegraphics[width=\columnwidth]{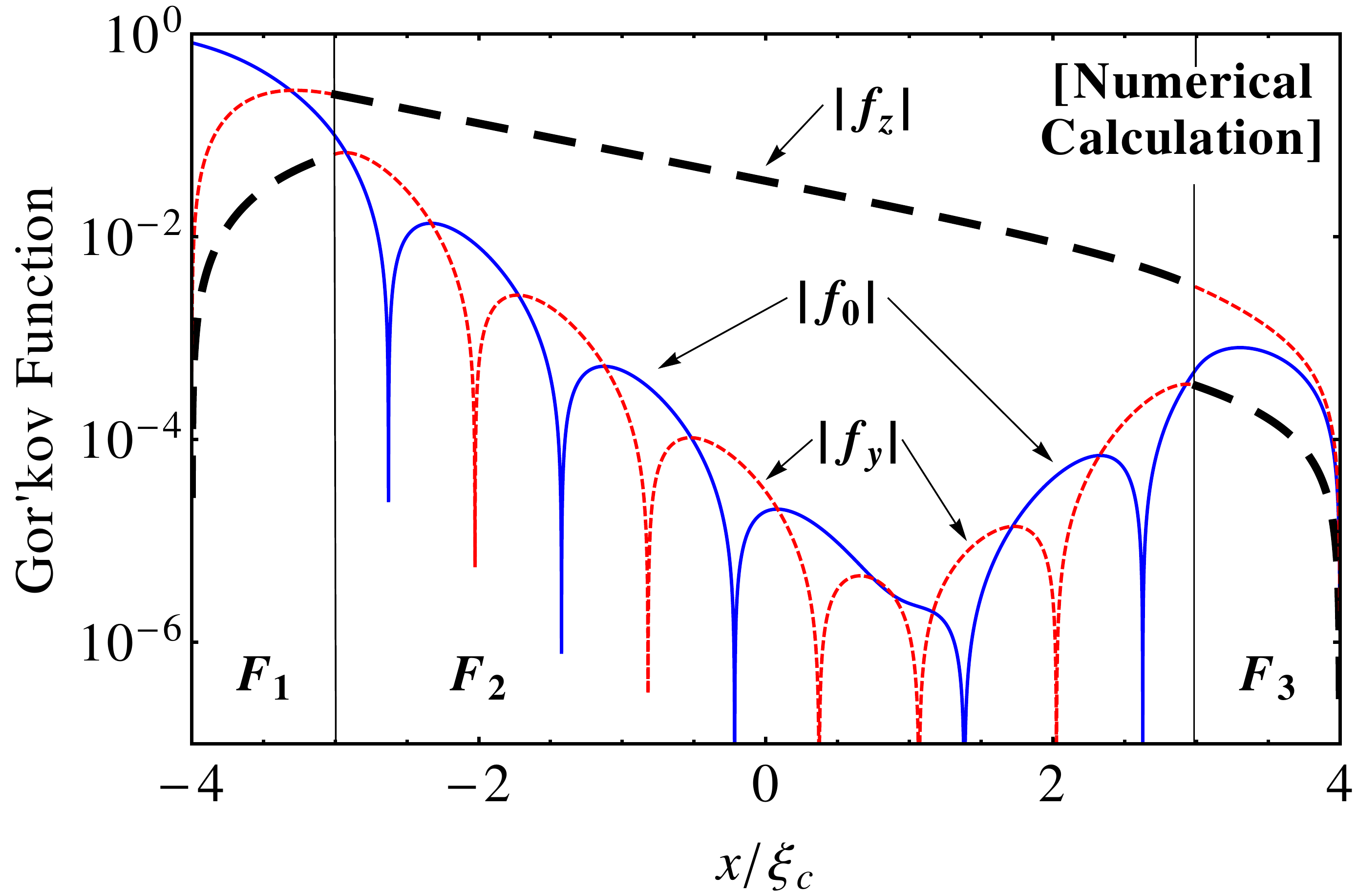}
\caption{\label{S3FS_Numerical}(color online) The Gor'kov functions for the spin valve SF$_1$F$_2$F$_3$S contributed from the left S only as obtained by solving the full Usadel equations numerically. The contribution from the right S is obtained by mirroring the curves in the figure about the $x=0$ vertical line (not shown).
Thick dashed lines (black) denote $m\neq0$ triplet components while solid lines (blue) denote singlets.  Dotted lines (red) show $m=0$ triplets. The line style and color denote the {\it symmetry} of pair correlations with respect to the local natural quantization axis and not the components of the Gor'kov functions; all functions $f_\alpha$ ($\alpha=0,y,z$) are {\it continuous} across the multilayer (see text). $\phi=\pi/2$, $h=(3,14,3)\pi T_c$, $d_F=(1,6,1)\xi_c$, $T = 0.4T_c$, $\omega_n=\omega_0$. Note that this figure is equivalent to Fig.~11 of Ref.~\onlinecite{bakerPRB16}, except that the thickness of F$_2$ in the latter reference is about double the thickness considered here, which explains why $|f_0|$ and $|f_y|$ are visible throughout the layer. 
}
\end{figure}

Figure~\ref{S3FS_Numerical} shows the Gor'kov functions for the spin valve when F$_2$ is oriented at a right angle with respect to F$_1$ and F$_3$ ($\phi=\pi/2$).
The line and color types used to represent the Gor'kov functions in this figure (and in Fig.~\ref{toymodelgorkov}) have a special meaning. They have been chosen to highlight the symmetry of pair correlations that appear in each F region.  Solid lines (blue) denote singlet pair correlations while dotted lines (red) highlight $m=0$ triplet correlations.  Thick, dashed lines (black) show $m=\pm1$ pair correlations.

It is important not to be confused by the linestyle and color code in Fig.~\ref{S3FS_Numerical} (and Fig.~\ref{toymodelgorkov}); we use the same convention as in Ref.~\onlinecite{bakerEPL14}. Each of the components $f_0,f_y,$ and $f_z$ is a {\it continuous} function of $x$ (as also seen in subsequent figures), in particular at each interface. But the symmetry of the correlations, as specified by the line style and color in Fig.~\ref{S3FS_Numerical}, changes at each rotation of the quantization axis. For example, $|f_z|$ is a dotted red line in F$_1$ and F$_3$ because it is the component parallel to the direction of the magnetization in these layers, Eq.~\eqref{3Fmag}, and thus represents the $m=0$ triplet state. The same component $|f_z|$ is a thick dashed black line in F$_2$ since the magnetization points in a direction perpendicular to $\hat{\mathbf{z}}$ and $f_z$ thus represents the $m\neq 0$ triplet correlations in F$_2$. The latter is the largest component in the center F (F\2) since the spin valve is constructed so that the $m=0$ contributions become negligible and only the $m=\pm1$ components contribute to the measured Josephson current.\cite{houzetPRB07}

\begin{figure}[b]
\includegraphics[width=\columnwidth]{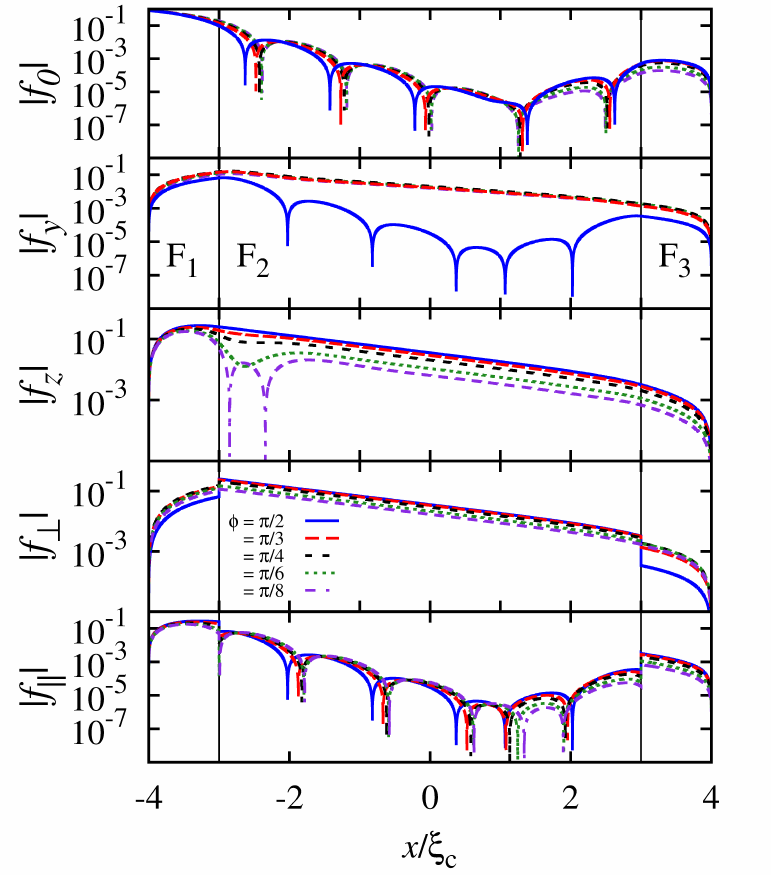}
\caption{\label{S3Fcantedgorkov}
(color online) Gor'kov functions $f_0$ and $\mathbf{f}$ in the SF$_1$F$_2$F$_3$ spin valve structure (S is not shown). From top to bottom, we display $f_0$ for the singlet, $f_y$ and $f_z$ for the triplet correlations. Also shown is the  triplet Gor'kov function, $\mathbf{f}$, decomposed in components perpendicular and parallel to the local magnetization direction.\cite{bakerPRB16}  One notices that the exponential decay of $m\neq0$ triplet components ($f_\perp$) is robust to canting. The $m=0$ components in $f_\parallel$ see their nodes shift slightly with the canting. Parameters used are  $h=(3,14,3)\pi T_c$, $T=0.4T_c$, and $d_F=(1,6,1)\xi_c$.
}
\end{figure}

Figure \ref{S3Fcantedgorkov} shows how the canting of the middle layer (F$_2$) magnetization affects the Gor'kov functions. We consider different values of $\phi$ in the range $[0, \pi/2]$ in Eq.~\eqref{3Fmag}. The case $\phi = \pi/2$ reproduces the result of Fig.~\ref{S3FS_Numerical} and the smallest angle for which the correlations are represented is $\phi = \pi/8$. The results would be unchanged had we considered the interval $[\pi/2,\pi]$. In Fig.~\ref{S3Fcantedgorkov}
each component $f_\alpha$ ($\alpha=0,y,z$) is presented on a separate plot and the line and color styles now distinguish different choices of the angle $\phi$. The breakdown into $m=0$ and $m\neq0$ triplets for each component is revealed by the oscillatory and smooth exponential decays, respectively.  One can differentiate more clearly the pair correlation contributions by representing the triplet Gor'kov function $\mathbf{f}$ in the rotating basis $\{\hat{\mathbf{x}}, \mathbf{e}_\perp(x),\mathbf{e}_\parallel(x)\}$ introduced in Ref.~\onlinecite{bakerPRB16}, rather than the Cartesian coordinate system of Fig.~\ref{cantedfigure}. The basis vectors $\mathbf{e}_\perp,\mathbf{e}_\parallel$ are perpendicular and parallel to the local magnetization $\mathbf{h}$ and thus depend on $\phi(x)$.
The $m\neq0$ components are in $f_\perp$ and $m=0$ components in $f_\parallel$.\cite{bakerPRB16}

A noteworthy feature of Fig.~\ref{S3Fcantedgorkov} is the presence of discontinuities in $f_\perp$ and $f_\parallel$. As stated earlier the components $f_\alpha$ ($\alpha=y,z$) are clearly continuous (see $|f_{0,y,z}|$ in Figs.~\ref{S3FS_Numerical}, \ref{S3Fcantedgorkov}), but the angle $\phi(x)$ of the magnetization is discontinuous across the interfaces, which results in the discontinuity of the components perpendicular and parallel to the magnetization. This is demonstrated by analyzing the relation between $f_{\perp,\parallel}$ and $f_\alpha$ ($\alpha=y,z$). Consider for example the behavior of $f_\perp(x) =  -\cos\phi(x)f_y(x) + \sin\phi(x)f_z(x)$ across the interface F$_L$F$_R$, with $\phi=0$ in the left F, and $0<\phi\ll \pi/2$ in the right F.  Using the continuity of $f_\alpha$ ($\alpha=y,z$) we have
\begin{eqnarray}
\left[f_\perp(x_+) - f_\perp(x_-)\right] 
&\approx& \left[ -\cos\phi(x_+) +\cos\phi(x_-)\right] f_y(x_i) \nonumber\\
&&+ \left[ \sin\phi(x_+) - \sin\phi(x_-) \right] f_z(x_i) \nonumber \\
&\approx& \phi f_z(x_i), \label{smallAngle}%\nonumber
\end{eqnarray}
where $x_\pm = x_i \pm \delta$, $x_i$ is the location of the interface and $0<\delta \ll 1$. This result shows that for continuous functions $f_{y,z}$, even at small misalignment of the magnetization at the interface F$_L$F$_R$ there is a jump of $f_\perp$ proportional to the mismatch angle $\phi$ and to the $m=0$ triplet component. At higher angle $\phi$ the conclusion remains but the relation is more complicated, to the point that even the sign of $f_\perp$ and $f_\parallel$ may change; this depends on the thicknesses of the F layers. The mismatch discussed here is not seen in the smooth continuous rotation of the magnetization $\phi(x)$ of an exchange spring or a helix.\cite{bakerPRB16}

A surprising result of Fig.~\ref{S3Fcantedgorkov} is seen in $f_\perp$. The canting may change the magnitude of the correlations by a factor of two or three, but essentially does not affect the order of magnitude of $m\neq0$ components in the 3F layer beyond $\phi \approx \pi/8$. This implies that no matter how the structure is canted in this range, the $m=\pm1$ components dominate.
The decrease in $f_\perp$ is evident as $\phi$ is decreased but this is a small change from the $\phi=\pi/2$ configuration on the logarithmic scale. Nevertheless, the small change in the Gor'kov functions is significant when considering the Josephson current.

\subsection{Effect of canting on the Josephson current %for small $d_{F_1}$
}\label{s:trilayerJc}

\begin{figure}
\includegraphics[width=\columnwidth]{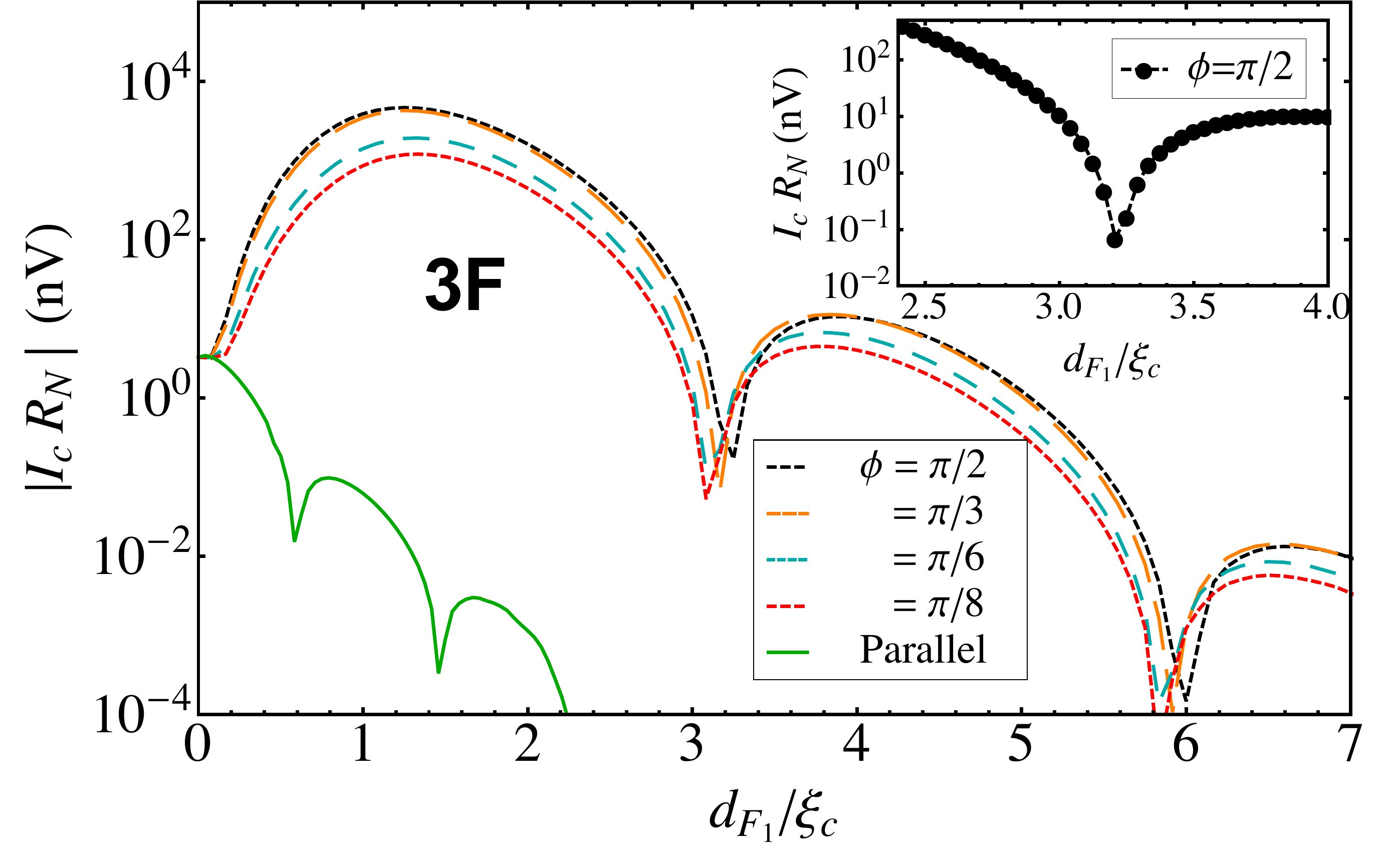}
\caption{\label{S3FS_current}(color online)
Josephson current as a function of the outer layers' thicknesses $d_{F_1}=d_{F_3}$ for a 3F structure at various cantings,$\phi$, of the central layer F$_2$.  A configuration where all F are parallel is also shown (lowest left solid green line). Usually, the nodes in the curve represent a sign change of the current (a crossing of the $I_c=0$ line); this is the case for the homogeneous configuration ($\phi=0$). However, in this figure the dips near $3.2\xi_c$ and $5.9\xi_c$ for $\phi>0$ are true minima {\it without} current reversal ($|I_c|>0$ for all $d_{F_1}$). The inset shows the {\it signed} current and demonstrates that $I_c>0$ through the dip; it is a DTM and not a node. This indicates a hidden change of sign in the Gor'kov functions (see text). $h=(3,14,3)\pi T_c$, $d_F=(d_{F_1},6,d_{F_1})\xi_c$, $T = 0.4T_c$.
}
\end{figure}
We briefly consider how the canting affects the Josephson current through the spin valve (Fig.~\ref{S3FS_current}) for typical thicknesses of $d_{F_1} \lesssim 2.5\xi_c$ considered in experiment and discuss features beyond $2.5\xi_c$ in Sec.~\ref{s:toymodel}. Even the small change in the Gor'kov functions is significant when considering the Josephson current, since this change is the reason for the ``hump like" structure seen in Fig.~\ref{S3FS_current} below $d_{F_1}/\xi_c \lesssim 3$, and also in Ref.~\onlinecite{houzetPRB07}. Note, however, that we consider only angles below $\phi=\pi/2$ and are thus not analyzing the $0-\pi$ transition discussed in Refs.~\onlinecite{houzetPRB07} and \onlinecite{bakerPRB16}. 

Comparing the curves for different canting angle $\phi$ in Fig.~\ref{S3FS_current} we note that increasing the canting notably increases the current flowing through the junction. The increase is most pronounced for small angles $\phi$. For example, at $d_{F_1} \simeq 2\xi_c$ the current increases several orders of magnitude as one goes from the homogeneous case ($\phi=0$) to even the smallest canting ($\pi/8$).  As one increases the canting further, say from $\phi=\pi/8$ to $\pi/2$ in Fig.~\ref{S3FS_current}, the growth of the current tends to level off; for example, the curves for $\phi = \pi/3$ and $\pi/2$ almost overlap. Although the growth of the current with increasing angle $\phi$ appears modest on the logarithmic scale, the current still increases by a factor of ten between the cases $\phi = \pi/8$ and $\pi/2$. The change in current as a function of canting angle away from $\phi = 0$ or $\pi$ is therefore large enough to be considered for applications.\cite{gingrichNP16}

Notwithstanding, we underline that the trilayer's $m\neq0$ components, and consequently the $I_c$, are quite robust to canting in the range of angles considered.  The largest variation in the current occurs for angles very close to $\phi = 0$. In Fig.~\ref{S3FS_current} the most drastic change in current occurs for $\phi \lesssim \pi/8$. 

It has been shown earlier that the critical current is a non-oscillating exponentially decaying function of the middle layer thickness $d_{F_2}$.\cite{khairePRL10,houzetPRB07} The same works also show that varying $d_{F_1}(=d_{F_3})$ leads to a non-monotonic current with a maximum. The latter behavior corresponds to the hump seen in Fig.~\ref{S3FS_current} for $d_{F_1}/\xi_c \lesssim 2.5$. This hump is similar to the one calculated in Fig.~2 of Ref.~\onlinecite{houzetPRB07}, except that the latter was represented on a linear scale and did therefore not analyze the behavior of the Josephson junction past $d_{F_1}/\xi_c \simeq 2.5$. It is also known that varying $d_{F_1}(=d_{F_3})$ affects the magnitude of the $m\neq0$ component in F$_2$, but current experimental studies consider the case where this thickness is small, of the order of $d_F=\xi_F$ to generate the maximal current.\cite{buzdinPRB11,khairePRL10,anwarPRB04,keizerN06,khairePRL10,robinsonS10,zhuPRL10,leksinPRL12,klosePRL12,wenEPL14,khasawnehSST11} Fig.~\ref{S3FS_current} extends the scope of these studies to reveal an interesting new feature, that we term ``depleted-triplet minima" (DTM) in the current seen at $d_{F_1} \simeq 3.2\xi_c$ and $\simeq 5.9 \xi_c$. This feature deserves special attention and the full discussion for $d_{F_1} \gtrsim 2.5\xi_c$ is postponed to Sec.~\ref{s:toymodel}.

\section{Pentalayer Spin Valves and canting}\label{s:pentalayer}
%\paragraph{Pentalayer Spin Valves}
In Ref.~\onlinecite{bakerEPL14}, we studied pair correlations in, and Josephson current through, a pentalayer with magnetizations of F\2 and F\4 perpendicular to F\1, F\3 and F\5 (see Fig.~\ref{cantedfigure}). In that paper we showed that a singlet component is present deep in the magnetic multilayer and is the origin of a Josephson current through the pentalayer. This challenged the common view that the $m=0$ components are only present near the interface between a singlet S and a F because they decay over the characteristic length $\xi_F$.  We introduced the cascade effect, the means by which all possible components are regenerated (with varying magnitude) at each rotation of the magnetization. Here we consider the same pentalayer to analyze how the canting angle $\phi$ of F\2 and F\4 affects pair correlations and the Josephson current through the multilayer.

\subsection{Effect of canting on pair correlations in the pentalayer}\label{ss:pentalayerPairs}

\begin{figure}
\includegraphics[width=\columnwidth]{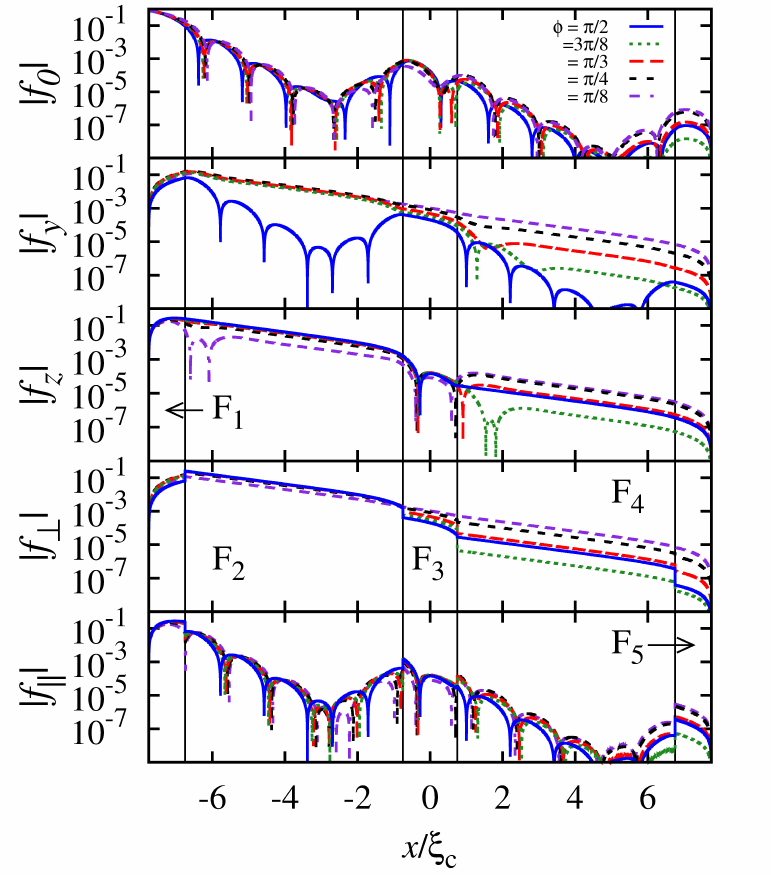}
\caption{\label{S5Fcantedgorkov}(color online)
Singlet ($f_0$) and triplet $f_{y,z}$ Gor'kov functions in the 5F for singlets leaking from a S located on the left of the pentalayer.  Also, shown are the pair correlations in the components perpendicular, $f_\perp$, and parallel, $f_\parallel$, to the local magnetization. These functions were calculated numerically using the techniques in Ref.~\onlinecite{bakerPRB16}. Each plot depicts one Gor'kov function for several angles.  The curve $\phi=\pi/2$ coincides with that shown in Ref.~\onlinecite{bakerEPL14}. Parameters: $h=(3,14,14,14,3)\pi T_c$, $T=0.4T_c$, and $d_F=(1,6,1.5,6,1)\xi_c$.}
\end{figure}
\begin{figure}
\includegraphics[width=\columnwidth]{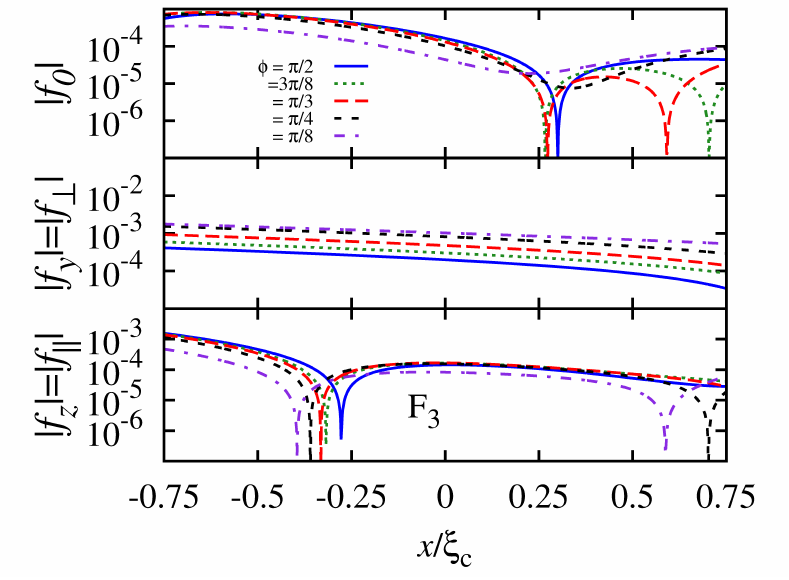}
\caption{\label{S5Fcantedgorkov_zoom}(color online) Same as Fig.~\ref{S5Fcantedgorkov} but focusing on the central layer F\3. Note the different ranges of the ordinates. In F$_3$ $f_y=f_\parallel$ and $f_z=f_\perp$ in F\3 since the magnetization of F$_3$ is along $\hat{\mathbf{z}}$. This identification does not apply to the canted layers F\2 and F\4 since the axes are not coincident with the magnetization's direction.}
\end{figure}
The pair correlations in the 5F pentalayer are shown as a function of the canting angle $\phi$ in F$_{2,4}$ in Figs.~\ref{S5Fcantedgorkov} and \ref{S5Fcantedgorkov_zoom}. The first feature to note is the opposite behavior of pair correlations in the central layer of the 5F and 3F spin valves (F\3 and F\2, respectively). In the 3F one starts with $m=0$ pair correlations in the homogeneous alignment and progresses to a domination of $m\neq0$ correlations, as one cants the central layer F\2 from parallel to perpendicular magnetization with respect to the outer layers. By contrast, the central layer F\3 of the pentalayer (Fig.~\ref{S5Fcantedgorkov_zoom}) has a dominant $m=0$ component in the {\it perpendicular} configuration (when F$_{2,4}$ have $\phi = \pi/2$) and progresses to one dominated by $m\neq0$ triplets close to the parallel alignment.\cite{bakerEPL14}

Similarly to the trilayer discussed in Sec.~\ref{s:trilayerJc}, the most drastic changes of the pair correlations in the pentalayer occur near $\phi = 0$ (and $\pi/2$ near SF). The growth of the $m\neq 0$ components is substantial and important for understanding the current as a function of canting angle as one lowers the value of $\phi$ from $\pi/2$ to $\pi/8$. But this component must collapse below the smallest angle since the $m\neq 0$ components are absent at $\phi=0$.

Both $|f_y|$  and $|f_z|$ of the pentalayer display the presence of $m\neq 0$ components, recognizable by the slow non-oscillatory decay of the correlations at various points. They also show $m=0$ oscillatory behavior at large angles in F$_{2,4}$ and  F$_{2,3,4}$, respectively. Not surprisingly, the behavior of $f_y$ and $f_z$, or $f_\perp$ and $f_\parallel$, as a function of canting is very similar in layer F\2 of the pentalayer and F$_2$ of the trilayer. In particular, $f_\perp$ is fairly robust to canting in F\2 of either structure.

Interesting are the correlations in layer F\4 of the 5F. The $m\neq0$ components in $f_\perp$ are much less robust to canting as in F$_2$ ; the component varies notably with $\phi$. Hence, the further from the SF (or FS) interface, the more sensitive $f_\perp$ (and to some extent $f_\parallel$) is to canting. As discussed below, this component is influenced by the $m=0$ components in the F\3 layer. The effect is opposite to that in F\2. With increasing $\phi$, one observes in Fig.~\ref{S5Fcantedgorkov} an increase (decrease) of $f_\perp$ in F\2 (F\4). As above (see Sec.~\ref{s:trilayer}), both $f_\perp$ and $f_\parallel$ components also have discontinuities at the interfaces where the magnetization is discontinuous.

Due to the cascade effect introduced in Ref.~\onlinecite{bakerEPL14}, the scalar singlet component $|f_0|$ is affected by the rotation of the magnetization, similarly to the 3F case of Fig.~\ref{S3Fcantedgorkov}. This is reflected in the fact that the curves do not exactly overlap for different angles $\phi$. Fig.~\ref{S5Fcantedgorkov_zoom} also highlights the resurgence of singlet components in the central layer F\3, due to the reverse FFLO effect.\cite{bakerEPL14} Finally, $f_0$ displays no oscillation and only has a minimum in F$_3$ at small, finite angles. The minimum monotonously deepens and moves towards the right interface as the angle increases. It crosses the $f_0=0$ line for $\phi > \pi/4$ as denoted by the two nodes. These nodes further move apart as one continues to increase the canting. When $\phi = \pi/2$ the minimum is located at or near the F$_3$F$_4$ interface; there is only one node in F$_3$. This means that $f_0$ has same sign at either boundaries of the F$_3$ layer for $0<\phi<\pi/2$, whereas the sign at either end is opposite for $\phi = \pi/2$, indicative of the drastic change in $f_0$ for angles close to $\phi=\pi/2$.

\subsection{Effect of canting on the Josephson current through the pentalayer}\label{ss:pentalayerJc}

\begin{figure}
\includegraphics[width=\columnwidth]{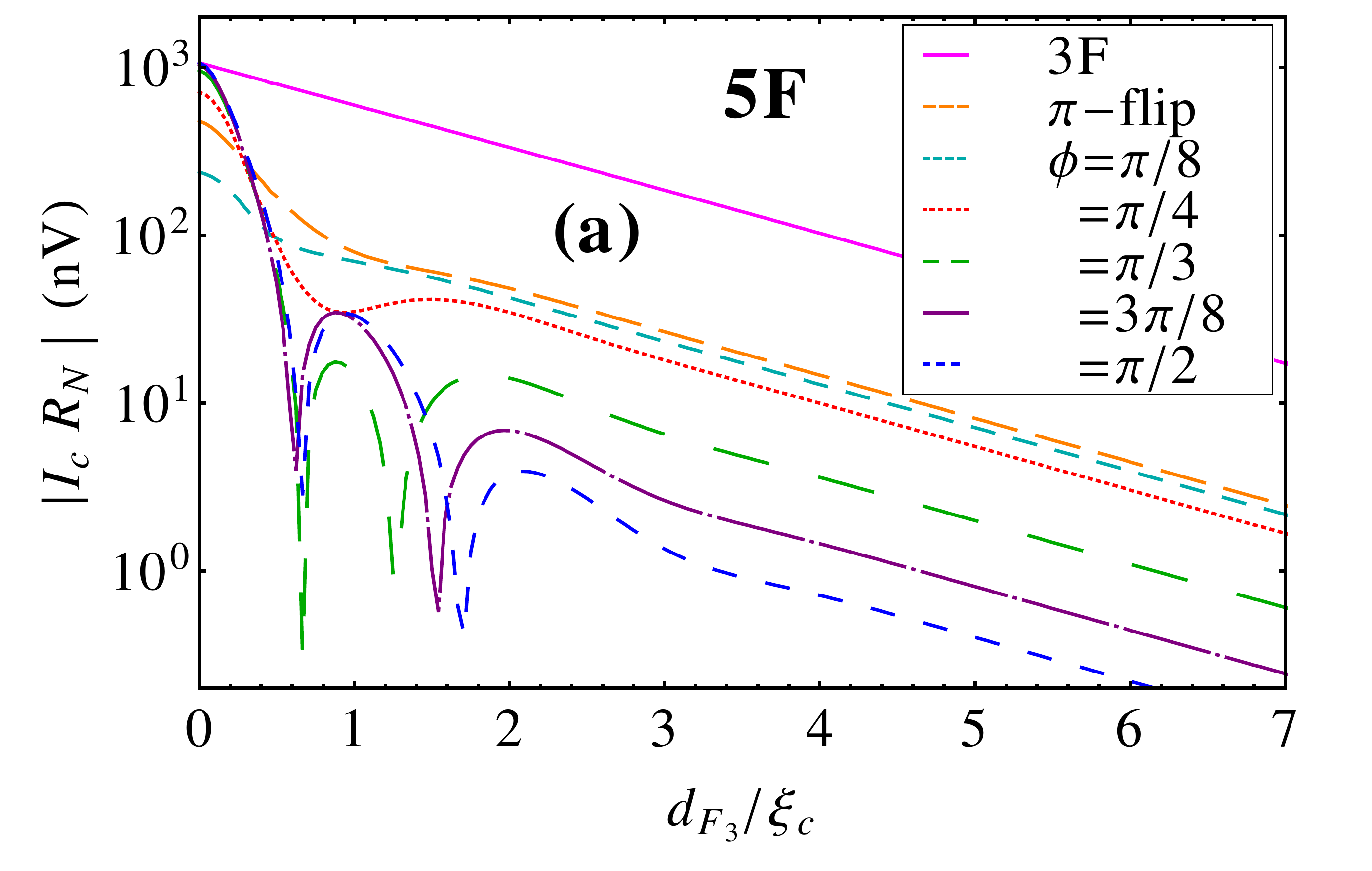}
\includegraphics[width=\columnwidth]{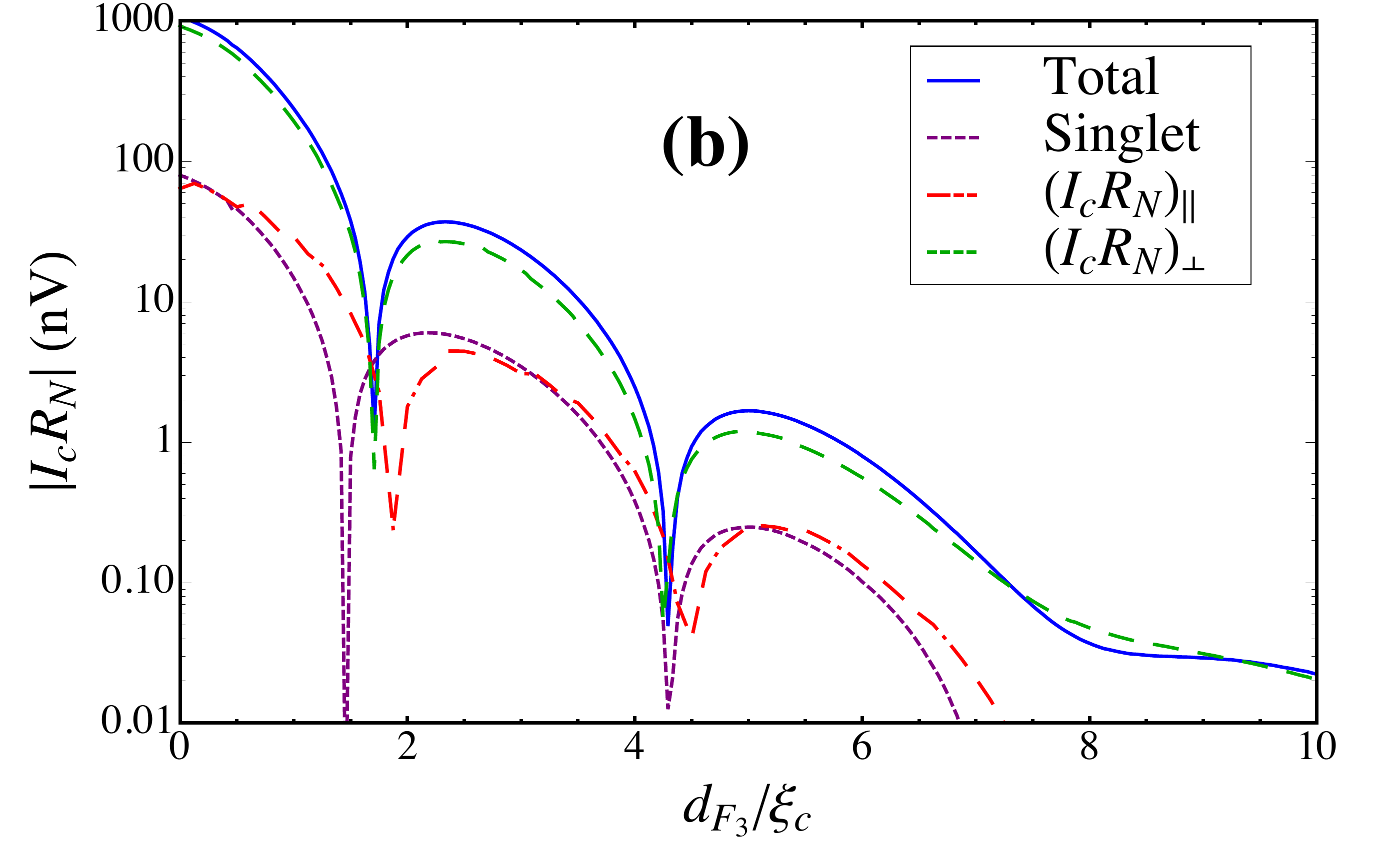}
\caption{(color online) \label{S5Fcurrent} Josephson current through a magnetic pentalayer as a function of the thickness of the central layer F\3. a) Current for different canting angles, $\phi$ (see Fig.~\ref{cantedfigure}).  Shown is also the ``$\pi-$flip" whereby each layer is misaligned by $\pi/4$ with respect to its neighbors, mimicking a Bloch domain wall (see full discussion of the continuous case in Ref.~\onlinecite{bakerPRB16}). The solid line labeled ``3F" is the current through a spin valve with $\phi=\pi/2$, with a smooth decay characteristic of the current entirely composed of $m\neq0$ correlations. b) Decomposition of the total Josephson current (solid line) for $\phi = \pi/2$ into the individual contributions of singlets correlations (dotted magenta), $m=0$ correlations, $(I_cR_N)_\parallel$ (dash-dotted red), and $m\neq 0$ correlations, $(I_cR_N)_\perp$ (dashed blue). Parameters: (a) $h=(3,14,14,14,3)\pi T_c$, $T=0.4T_c$, and $d_F=(1,6,d_{F_3},6,1)\xi_c$. (b) As in (a) but $\phi=\pi/2$ and $h_{F_3}=3\pi T_c$.
}
\end{figure}

Figure~\ref{S5Fcurrent}a shows the Josephson current flowing through the 5F layer as a function of the central layer thickness $d_{F_3}$ for different canting angles $\phi$ of layers F\2 and F\4 with respect to the $\hat{\mathbf{z}}$ axis ($\phi_{F_2} = \phi_{F_4}=\phi$; see Fig.~\ref{cantedfigure}). We first focus on the general features of the current as a function of $d_{F_3}$ at the fixed value $\phi = \pi/2$. As discussed in Ref.~\onlinecite{bakerEPL14}, the current displays a characteristic $0-\pi$ oscillation at thicknesses $d_{F_3} \lesssim 2.5\xi_c$, recovering the physics of $m=0$ correlations deep in the magnetic layer.\cite{bakerEPL14} The oscillation is revealed by the presence of two nodes seen in the lower dashed blue line of Fig.~\ref{S5Fcurrent}a.  Beyond $d_{F_3} \sim 2.5\xi_c$ the $m=\pm1$ correlations start dominating, leading to the monotonic exponential decay over the longer length scale $\xi_N$.

As one changes the canting in the interval $\phi\in [0,\pi/2)$ one identifies three regimes for the current in Fig.~\ref{S5Fcurrent}a: $d_{F_3} \lesssim \xi_c/2$, $\xi_c/2 \lesssim d_{F_3} \lesssim 2.5\xi_c$ and $d_{F_3} \gtrsim 2.5 \xi_c$.

For $d_{F_3}\lesssim0.5\xi_c$ the current rapidly increases as a function of canting angle $\phi$ towards the value of the 3F  perpendicular configuration (where the magnetization in F\2 is perpendicular to F\1 and F\3). The increased current is due to the generation of a stronger $f_\perp$ component with larger $\phi$.  

Fig.~\ref{S5Fcantedgorkov_zoom} can be used to understand the behavior of the Gor'kov function components for these smaller thicknesses.  The Gor'kov functions look very similar to this figure if the right edge  is moved the appropriate distance from the left. For example, if a layer of $d_{F_3}=\xi_c$ is required, we could simply cutoff Fig.~\ref{S5Fcantedgorkov_zoom} a distance $\xi_c$ away from the F$_2$F$_3$ interface.  In this way, the parallel components are seen to dominate over the perpendicular components.

In the opposite regime of large central layer thicknesses $d_{F_3} \gtrsim 2.5 \xi_c$ the current in Fig.~\ref{S5Fcurrent}a displays a behavior that may appear counterintuitive at first. The current is monotonously decaying with $d_{F_3}$ but {\it increases} in magnitude overall with {\it decreasing} angles. Again, this can be understood from the Gor'kov functions. The general trend of the $m\neq0$ components for large $d_{F_3}$ are the same as those in a long trilayer, already shown in Fig.~\ref{S3FS_Numerical} (but labeling F$_i$ with $i_\mathrm{ 3F} \leftrightarrow (i+1)_\mathrm{5F}$ for $i=1,2,3$). In the pentalayer, the $m=0$ correlations generated from the cascade effect\cite{bakerEPL14} at the F$_2$F$_3$ interface decay quickly in F$_3$ and the $m \neq 0$ components dominate, as is the case of the central layer in the trilayer. The study of the latter would lead one to think that the $m \neq 0$ components increase with canting leading to an increase of the current. However, the opposite is observed for the pentalayer in Fig.~\ref{S5Fcurrent}a. The reason is that in the 5F the $m\neq 0$  components are already generated in F$_2$ and they increase with canting in that layer.  Thus, when entering the central layer F$_3$, more of these components will be transformed into $m=0$ components, implying a {\it decrease} of $f_\perp$ with increased canting in F$_3$.  Since the $m=0$ components are not contributing much to the current in this configuration, the decrease of $m \neq 0$ components in F$_3$ leads to lower current $I_c$, as observed for $d_{F_3}\gtrsim 2.5\xi_c$ in Fig.~\ref{S5Fcurrent}a.

Finally, for $0.5 \lesssim d_{F_3}\lesssim 2.5 \xi_c$, the sign of the current in Fig.~\ref{S5Fcurrent}a changes as one increases the canting angle. The structure is undergoing a $0-\pi$ transition of the Houzet-Buzdin type since in this regime the dominant contribution to the current are $m\neq0$ components.\cite{houzetPRB07,bakerPRB16} This type of $0-\pi$ transition was originally proposed in trilayers while changing the magnetization direction in F\1 and F\3 with one orientation changing.\cite{houzetPRB07} However, there is an important difference in the pentalayer. While in the trilayer the $m=0$ components play no role whatsoever, in the pentalayer (5F) they are controlling the sign of the $f_\perp$ components and thus the direction of the current.  To understand this point we consider the pair correlations in Fig.~\ref{S5Fcantedgorkov_zoom} (or Fig.~\ref{S5Fcantedgorkov}). We first note that $f_\perp$ does not change sign in F$_3$. Yet, $f_\perp$ has opposite sign in F\2 and F\4 for $\phi=\pi/2$. The node leading to this sign change is in the continuous function $f_z$ and appears in F\3. While $f_z = f_\perp$ in both F\2 and F\4 (where $\phi=\pi/2$), we have $f_z = f_\parallel$ in F\3 (where $\phi=0$). Thus, the correlations represented by $f_\perp$ in F$_2$ are continued by the $f_\parallel$ curve in F\3 before continuing back into the $f_\perp$ curve in F\4. The node of  $f_\parallel = f_z$ in F\3 implies a sign change of $f_\perp$ in F\4. Thus, in the intermediate thickness regime the $m=0$ components in the central layer F\3 determine the sign of $f_\perp$ across the pentalayer and the existence of a $0-\pi$ transition. In the pentalayer the $m=0$ components control this Houzet-Buzdin transition.

To corroborate this statement we point out that at lower canting angles, the nodes of the $m=0$ components disappear in Fig.~\ref{S5Fcantedgorkov_zoom}, concurrent with the disappearance of $0-\pi$ transitions in the current; as the canting angle decreases, the nodes on either side of $d_{F_3}\sim \xi_c$ move towards each other and disappear for $\phi \lesssim \pi/3$. This indicates that the minimum of the current has shifted above the $I_c=0$ axis and there are no longer $0-\pi$ transitions. The smooth decay of the critical current for fixed $\phi<\pi/3$ indicates that there are no $m=0$ correlations of noticeable strength to the current.

To complete this discussion, note that the remarks made in Sec.~\ref{ss:pentalayerPairs} about pair correlations for small canting angles ($\phi<\pi/8$) transfer to the current. As the canting decreases to $\phi=0$ it must tend towards the same current found in a homogeneous configuration, where there are no $m\neq0$ components and the current is much lower for this thickness. Hence, the changes in the current are most drastic close to $\phi = 0$. This does not diminish the observation that the current through the pentalayer is much more susceptible to canting than the more robust current through the trilayer.

Figure~\ref{S5Fcurrent}b ($\phi = \pi/2$, and weaker magnetization $h_{F_3}$) is similar to the result presented in Fig.~4 of Ref.~\onlinecite{bakerEPL14} but disentangles the contribution of parallel ($m=0$) and perpendicular ($m\neq 0$) components of the Gor'kov functions to the Josephson current using the techniques of Ref.~\onlinecite{bakerPRB16}. While in the latter reference we disentangled the contributions for various domain wall twists at fixed thickness of the hybrid structure, Fig.~\ref{S5Fcurrent}b considers a fixed canting, and varies the thickness of the central layer $d_{F_3}$. We first note that both the parallel and perpendicular components of the current undergo a sign change in the intermediate regime (this regime spans a larger $d_{F_3}/\xi_c$ range than in Fig.~\ref{S5Fcurrent}a because we chose $h_{F_3}$ weaker). As discussed above, the sign change of the perpendicular component is determined by the $m=0$ correlations and the perpendicular component remains dominant at all thicknesses. This is revealed by two features in the figure. The magnitude $(I_cR_N)_\perp$ is similar to that of the total current. Furthermore, the nodes of this component of the current almost exactly coincides with the nodes of the total current, while the $m=0$ contributions lead to nodes that are slightly shifted. These features indicate that the $m\neq 0$ correlations determine the behavior of the current in this intermediate regime.

It would be of great interest to perform a measurement of the Josephson critical current through a pentalayer heterostructure as a function of the middle layer thickness F\3. The variation of the current with decreasing angle would provide direct evidence of the $m=0$ and $m\neq 0$ Gor'kov functions contributions. The results could be compared to the trilayer case to show that, while $m=\pm1$ components are robust to canting in the trilayer, the five layer system tunes in $m=0$ components.

Finally, we point out one more difference between the trilayer and pentalayer. As shown in Figs.~\ref{S5Fcantedgorkov_zoom} and \ref{S5Fcurrent} the Gor'kov functions and the Josephson critical current, respectively,  do not display the same monotonous behaviour as a function of $\phi$ as in the trilayer case. A continuous increase of the angle does not necessarily translate into a continuous increase or decrease of correlations and current. For example, the $\phi = 3\pi/8$ curve in Fig.~\ref{S5Fcantedgorkov} is below the $\pi/2$ line.  This is particularly visible for $f_z$ and $f_\perp$ in the F$_4$ layer.  Similarly, the current may be a non-monotonous function of angle $\phi$ when the thickness of F$_3$ is near a node of the current (a $0-\pi$ transition). As the angle decreases from $\phi=\pi/2$ in Fig.~\ref{S5Fcurrent}a, the second node shifts to lower values of $d_{F_3}$. As a result, for example at fixed thickness $d_{F_3} \simeq 1.5\xi_c$ the current at $\phi=3\pi/8$ is lower than at $\pi/2$. The pentalayer thus displays a much richer physics that could be advantageous for applications.

\section{Depleted-Triplet Minima in the Josephson Current of a Spin Valve}\label{s:toymodel}

A thorough study of the effects of canting on the magnetic structure has been conducted in the previous sections for the trilayer and pentalayer spin valve systems.  One remarkable effect that we reserved for this section is the presence of the depleted-triplet minima seen in Fig.~\ref{S3FS_current} near $d_{F_1} \sim 3\xi_c$ and $6\xi_c$. Although they seem to be similar to the dips found in all other figures, they do not signify a sign change of the current and have a different origin. This feature of the current through a trilayer has not been previously discussed in the literature and occurs at arbitrary, but fixed canting as one varies the thickness of the external magnetic layers F$_1$ and F$_3$ in the trilayer. The phenomenon is found when one extends the study beyond the range of thicknesses considered so far experimentally and theoretically, that is, for $d_{F_1}=d_{F_3} \gtrsim 2.5\xi_c$ for the parameters of Fig.~\ref{S3FS_current}. As noted in the inset of that figure the dips do not cross the $I_c=0$ line, meaning they are not indicative of $0-\pi$ transitions on the logarithmic scale but are true minima. Because the figures for the Josephson current generally plot the absolute value of the current the feature is not readily identifiable in the representation of Fig.~\ref{S3FS_current}. This section is focused on revealing why this feature appears.

As discussed above, previous work demonstrated that a $0-\pi$ transition can be generated in the trilayer structure by increasing the angle of one of F\1 and F\3 beyond $\phi=\pi/2$.\cite{houzetPRB07,khairePRL10} Alternative procedures to generate a $0-\pi$ transition are either to set $\phi=0$ (homogeneous case) and vary the thickness $d_{F_1} = d_{F_3}$ or to fix the canting angle to a finite value but vary the thickness of one of the two layers only ($d_{F_1}$ or $d_{F_3}$), which will be shown in Sec.~\ref{ss:trilayerJctoy} (see Fig.~\ref{3Fcurrentsfocus}, curve E).  All these current reversal transitions were shown to relate to the change of relative sign of the Gor'kov functions generated from the left and the right S.\cite{houzetPRB07,bakerPRB16} This is not the situation encountered with the DTMs of Fig.~\ref{S3FS_current} since the current does not change direction.

Two ingredients lead to the presence of the DTMs in Fig.~\ref{S3FS_current}. First, as the thickness increases through $d^{\rm DTM}_{F_1}$ ($\approx3.2\xi_c$) the sign of the $m=\pm 1$ Gor'kov functions from either S changes {\it at the same time}. In the case of Fig.~\ref{S3FS_current} the Gor'kov function from the left and right superconductor are negative at the SF$_1$ and F$_3$S interfaces when $d_F \leq d^{\rm DTM}_{F_1}$, while {\it both} are positive when $d_F > d^{\rm DTM}_{F_1}$. The simultaneous change of sign is due to the symmetric treatment of F$_1$ and F$_3$; as a result, no node should be observed. By contrast, the aforementioned procedures to generate the $0-\pi$ transition rely on an asymmetric treatment of F\1 and F\3.

A second ingredient is necessary to explain the DTM. Inspection reveals that at the thickness $d^{\rm DTM}_{F_1}$ the Matsubara frequency-averaged position of the nodes of $m=0$ Gor'kov functions coincides with the interface. Since the $m=0$ components are nearly zero at the interface the $m\neq 0$ pair correlations amplitudes are depleted in F\2. Both factors lead to a minimum in the Josephson current rather than a $0-\pi$ transition.

Demonstrating unequivocally that the DTM feature near $d_{F_1}/\xi_c \sim 3$ in Fig.~\ref{S3FS_current} is not a $0-\pi$ transition is challenging for the numerical techniques used here as it would require to calculate the Gor'kov functions for all Matsubara frequencies with great precision close to the DTM. It is much more convincing to reveal the effect by constructing a toy model of the trilayer from an analytic solution\cite{bakerEPL14} to show that the sign of the Gor'kov function as contributed from the left and the right change together, hence avoiding a sign change in $I_c$.

We construct the toy model following simple rules to provide a clearer picture of the full numerical calculation presented in Sec.~\ref{ss:pentalayerJc}. The purpose of the toy model is not to emulate the physics entirely but to bring to light specific behaviors.  We will specify which features are not reproduced by the toy model.

\subsection{Pair correlations in the toy model}\label{s:trilayerPairstoy}

The main simplification of the toy model consists in setting the magnetization to zero artificially for homogeneous layers with magnetization  perpendicular to the outer layers. This approximation results from the known fact that the component perpendicular to $\mathbf{h}$ propagates as though it were in a normal metal.\cite{buzdinRMP05,bakerPRB16} Hence, choosing the magnetization direction of F\1 along $\mathbf{\hat{z}}$, we set $h=0$ in F\2.  It is important to realize that this model does not simply describe an FNF system where a normal metal (N) is sandwiched between two Fs. Rather, our toy model is equivalent to an FNF with spin active FN and FN interfaces to generate the $m\neq 0$ triplet components in F$_2$. The solution of the Usadel equation for the toy model is best described in terms of the alternate trigonometric parameterization used in Refs.~\onlinecite{zaikinZP81,belzigS&M99,houzetPRB05,vasenkoPRB08,faurePRB06,bakerJSNM12,bakerEPL14}. The Gor'kov function in this spin valve configuration as contributed from the left S (indexed by $L$) takes the form of a piecewise function
\begin{equation}\label{LRTCgorkovfct}
\hspace{-6cm}\mathcal{F}_L(x,h,\theta_0)=
\end{equation}
\begin{equation}
\nonumber
\begin{cases}
f_0+if_z=\sin\theta(x,h,\theta_B), & x\in F_1 \\[2ex]
f_0+if_y=\sin\theta(x,h,\mathrm{Re}[\theta(-d_\mathrm{F_2}/2)])\\
\mathrm{and}\quad if_z=\sin\theta(x,0,i\mathrm{Im}[\theta(-d_\mathrm{F_2}/2)]), & x\in F_2\\[2ex]
f_0+if_z=\sin\theta(x,h,i\mathrm{Im}[\theta(d_\mathrm{F_2}/2)]), & x \in F_3,
\end{cases}
\end{equation}
where $\theta(x,h,\theta_0)$ is the complex function parametrizing the Gor'kov functions, $x$ is defined as in Fig.~\ref{cantedfigure}, $h$ is the magnitude of the magnetization in each layer, $\theta_0$ is the boundary value of $\theta$ for the given F layer at the left edge and determined by the value of $\theta$ in the adjacent left layer, $\theta_B$ is the bulk value in S defined above Eq.~\eqref{eq: gorkovadd}.
The real part of $\theta$, $\mathrm{Re}[\theta] \equiv \vartheta$, is defined in the parametrization of Sec.~\ref{s:method} (see also Ref.~\onlinecite{bakerEPL14,bakerPRB16}). The imaginary part $\mathrm{Im}[\theta]$ determines the functions $M_0$ and $\mathbf{M}$ in Refs.~\onlinecite{parametrization,bakerPRB16}.
The function describing the contribution from the right S ($\mathcal{F}_R$) is found by setting $x\rightarrow-x$
and F$_1\leftrightarrow$F\3.

The imaginary parts of the Gor'kov function $\mathcal{F}$ ($f_y$ in F\2 or $f_z$ in F$_{1,3}$) denote $m=0$ triplet correlations and are the components parallel to the magnetization; these are the signature of the FFLO effect.\cite{larkinJETP65,fuldePRL64}

In the analytic toy model proposed here only the values of the functions are matched at the interface. Alternatively, one could construct a similar model of the system by ensuring the derivatives match. The toy model can be interpreted as a full calculation with different interface transparencies.\cite{vasenkoPRB08} This is one contrasting feature of the toy model since the full numerical calculations presented in this paper are performed for transparent interfaces; both the value and derivatives of the Gor'kov functions match across the interfaces.

\begin{figure}[t]
\includegraphics[width=\columnwidth]{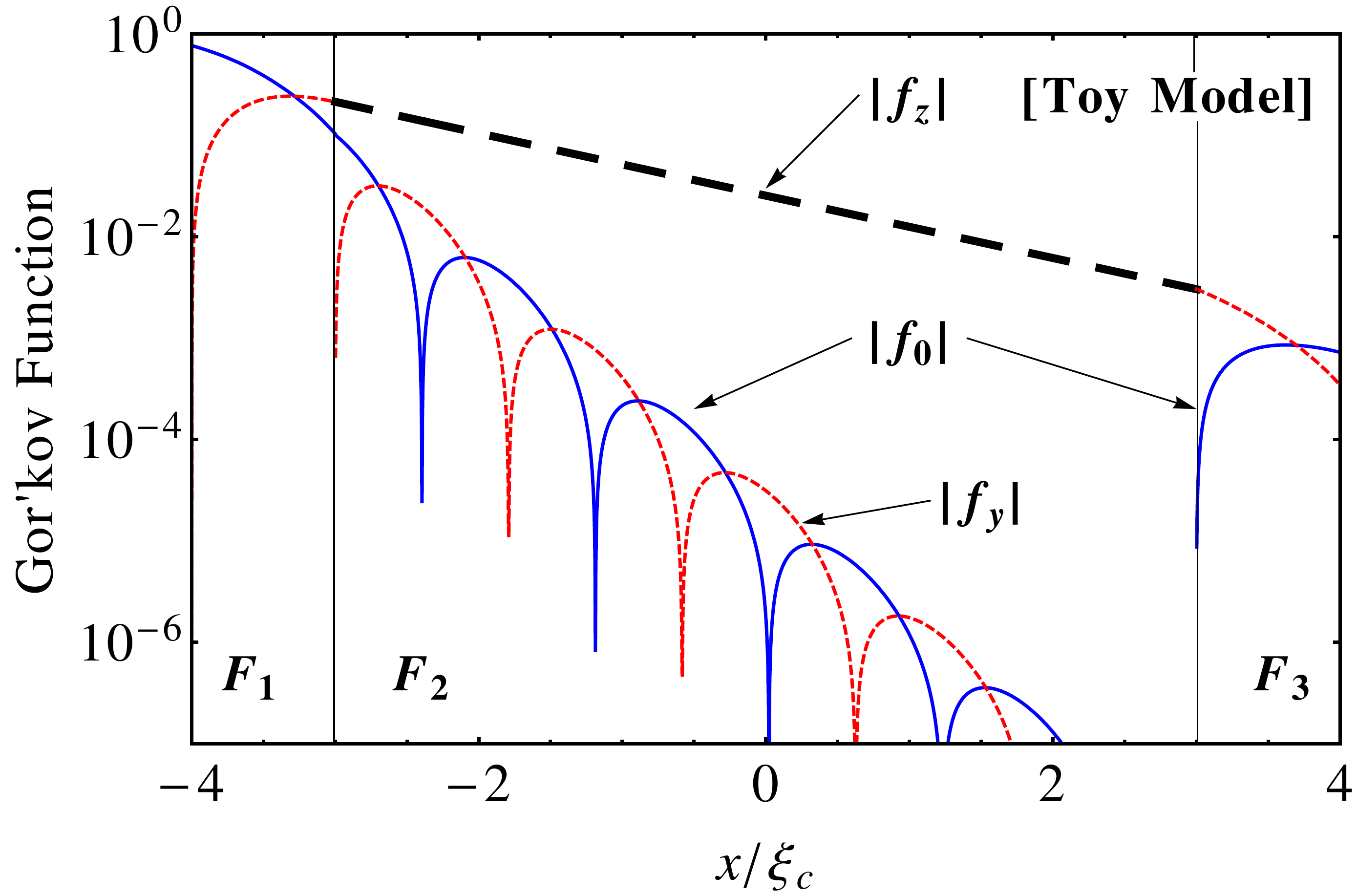}
\includegraphics[width=\columnwidth]{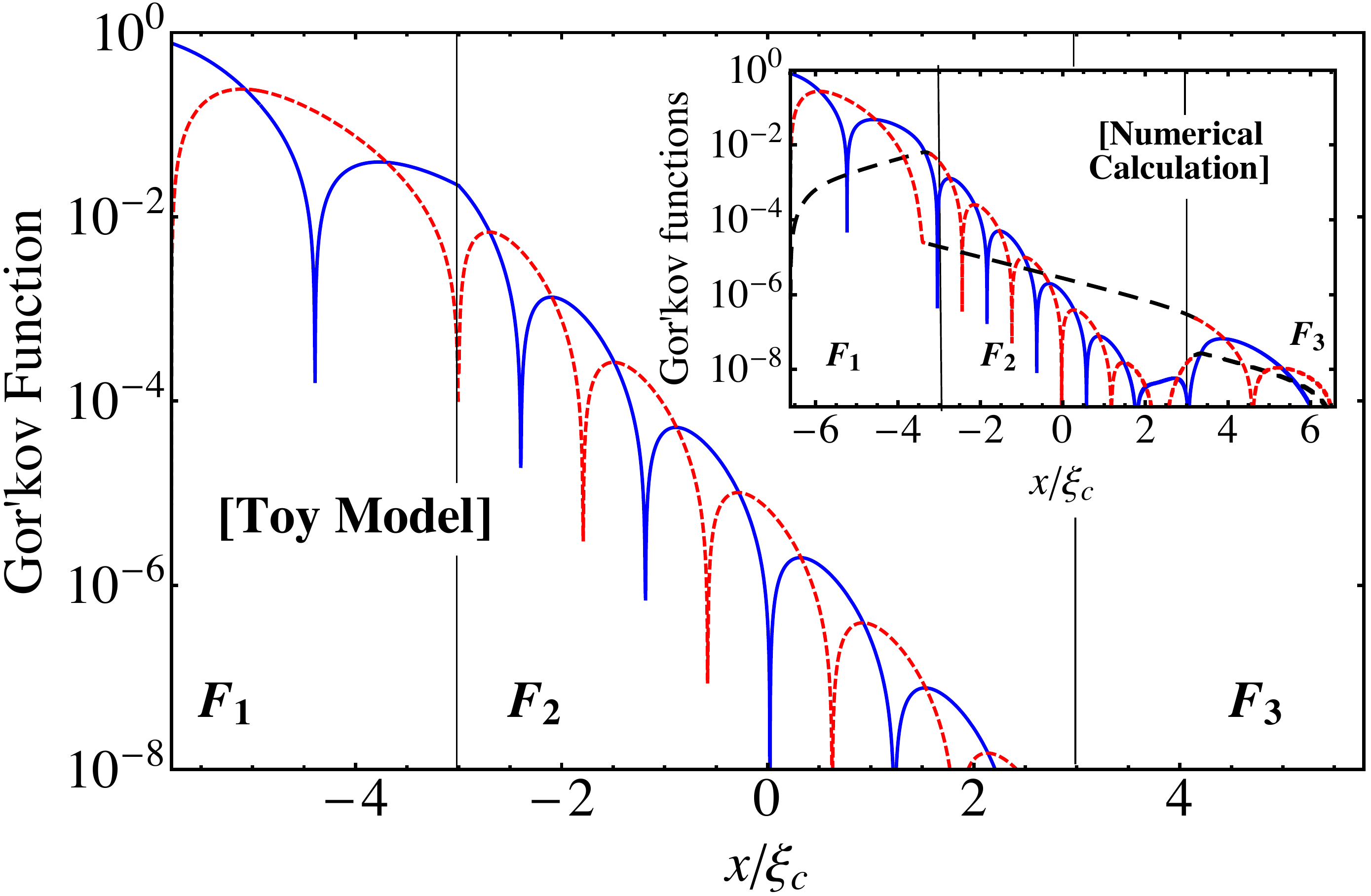}
\caption{\label{toymodelgorkov} (color online) A toy model calculation, Eq.~\eqref{LRTCgorkovfct}, of pair correlations in a trilayer spin valve structure for two different thicknesses of F$_{1,3}$ ($d_{F_1}=d_{F_3}=\xi_c$).  The line style and color code are explained in Fig.~\ref{S3FS_Numerical} and refer to the {\it symmetry} of pair correlations. (a) Solution of the toy model for the same parameters as Fig.~\ref{S3FS_Numerical}, allowing to compare the toy model with the full numerical solution (see text). (b) Calculation for thicknesses $d_{F_{1,3}} \simeq 2.79\xi_c$ ($d_{F_{1,3}} \simeq 3.21\xi_c$ for the inset) corresponding to the location of the minimum of the DTM in Fig.~\ref{3Fcurrentsfocus} (and the first DTM in Fig.~\ref{S3FS_current} for the inset). The critical feature seen in Figure b is that a node of $|f_y|$ exactly coincides with the F$_1$F$_2$ interface (see text). Both figures have magnetization configuration $(\phi_{\mathrm{F}_1},\phi_{\mathrm{F}_2},\phi_{\mathrm{F}_3})=(0,\pi/2,0)$. Parameters are  $h=(3,14,3)\pi T_c$, $d_{F_2}=6\xi_c$, $T = 0.4T_c$, $\omega_n=\omega_0$.
}
\end{figure}

Figure \ref{toymodelgorkov}a displays $\mathcal{F}_L$, the solution of the Usadel equations with Eq.~\eqref{LRTCgorkovfct}. As in Fig.~\ref{S3FS_Numerical} (see discussion in Sec.~\ref{s:trilayerPairs}), we chose the line style and color of the curves in each layer to identify the pair correlations symmetries $\ket{s,m}$ rather than the components ($f_\alpha$, $\alpha=0,y,z$); the Gor'kov functions are continuous across interfaces. Comparing Fig.~\ref{toymodelgorkov}a with Fig.~\ref{S3FS_Numerical} allows to identify the components present in the full numerical calculation that are absent in the toy model. For example, the $m=0$ triplet components arise in F\pt{$_1$} (dotted red line) and generate $m \neq 0$ triplet components in  F\2 (dashed black line). Hence, the toy model  accurately provides an $m\neq 0$ triplet component along $\mathbf{\hat{y}}$ in F\2.\cite{bakerEPL14}   On the other hand, the amplitude of the $m=0$ components $f_0$ and $f_y$ in F\2 generated at the F\2F\3 interface are not seen in Fig.~\ref{toymodelgorkov}a since they are orders of magnitude smaller. For this same reason $m\neq 0$ components are absent in F\3, while $m=0$ are of substantial magnitude. It is noteworthy that even in the toy model the singlet reappears in the F\3 layer.\cite{bakerEPL14} This difference between the two cases owes to the absence of ``back diffusion" of pair correlations in the toy model. In Fig.~\ref{toymodelgorkov}a only pairs diffusing from the left S to the right appear; the toy model neglects components in each layer that result from reflections at the interfaces.
Figs.~\ref{S3FS_Numerical} and \ref{toymodelgorkov}a demonstrate that the main feature of the spin valve is reproduced by the toy model, since both reveal the $m\neq0$ component as the dominant pair correlation.

Figure \ref{toymodelgorkov}b reveals the essential feature that explains why there is a DTM in the Josephson current, Figs.~\ref{S3FS_current} and \ref{3Fcurrentsfocus}. At $d_{F_1}^{\mathrm{DTM}}$ the first node of the $m=0$ component (dotted red) coincides with the F\1F\2 interface. This minimizes the formation of $m \neq 0$ correlations ($|f_z|$) at that interface and thus depletes the pair correlations contributing to the Josephson current (see next section). As $d_{F_1}$ varies through $d_{F_1}^{\mathrm{DTM}}$ the node approaches and passes the interface, resulting in a minimum at $d_{F_1}^{\mathrm{DTM}}$ of $m\neq 0$ pair correlations in F$_2$.  
The small $m\neq 0$ component (dashed black line) seen in the inset of Fig.~\ref{toymodelgorkov}b is orders of magnitude smaller than in Fig.~\ref{toymodelgorkov}a. This residual component comes from the inability to choose a thickness of F$_{1,3}$ with enough numerical accuracy to have the node of the $m=0$ component exactly coincide with the position of the node. This underlines the advantage of the toy model.

\subsection{The Josephson current in the toy model}\label{ss:trilayerJctoy}

\begin{figure}
\begin{center}
\includegraphics[width=\columnwidth]{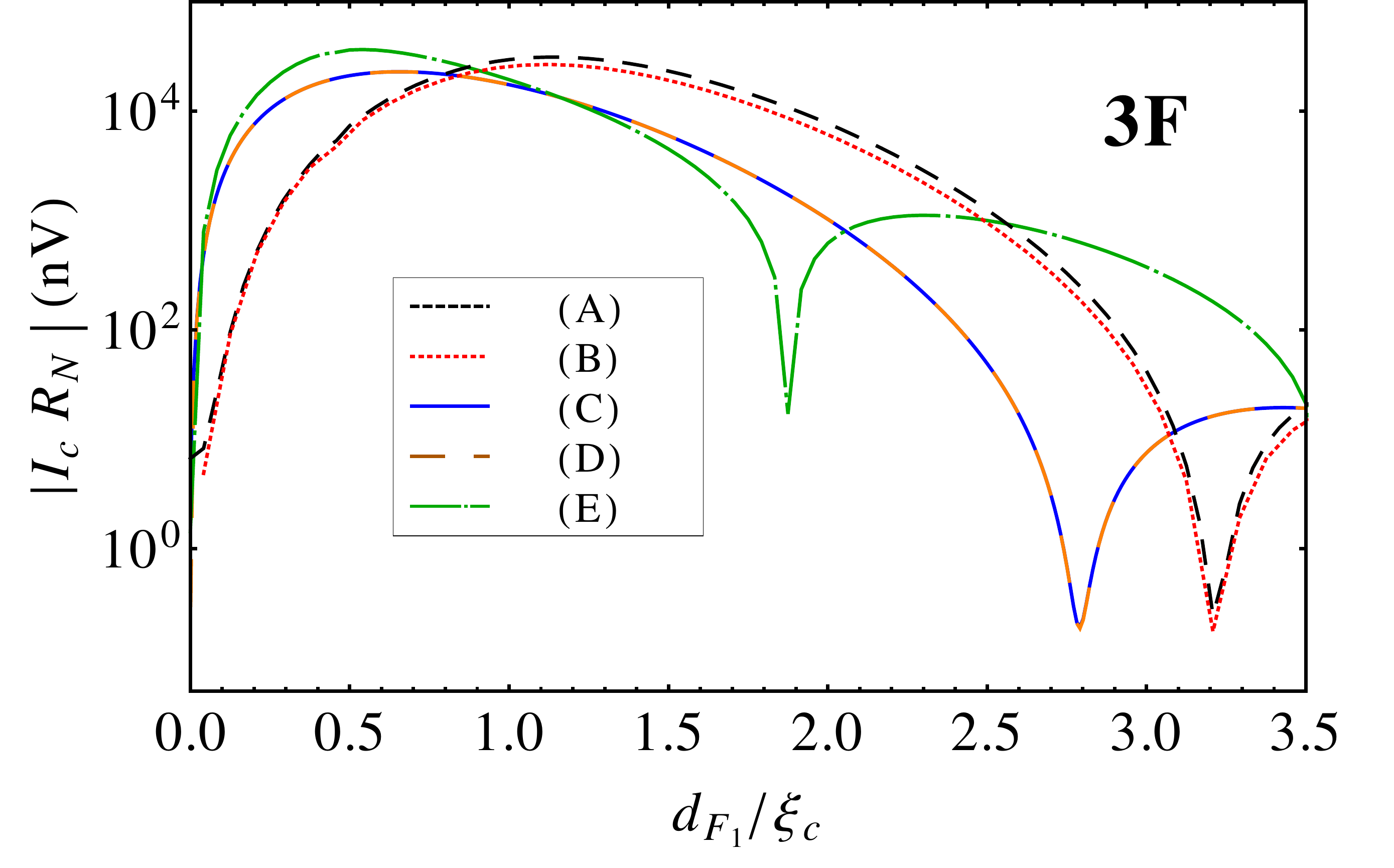}
\end{center}
\caption{\label{3Fcurrentsfocus} (color online) Critical current through a trilayer spin-valve in the perpendicular configuration as a function of the thickness $d_{F_1}$ ($=d_{F_3}$ for curves A-D). (A) Full numerical calculation; same as the curve for $\phi=\pi/2$ in Fig.~\ref{S3FS_current}. (B) Same as (A) but neglecting the singlet contribution (only  triplet current).  (C) and (D): Same as (A) and (B), respectively, but using the toy model, Eq.~(\ref{LRTCgorkovfct}). (E) Full numerical calculation for the case when $d_{F_3}=\xi_c$ for all values of $d_{F_1}$ (this is the only case where $d_{F_1} \neq d_{F_3}$ in the figure).  (A) and (B), and (C) and (D) essentially overlap, indicating that the cascading singlets only minimally influence the current. As the text explains, the toy model shows that the dip in the current, the DTM in (A)-(D), is not a $0-\pi$ transition but a finite minimum at $d_{F_1}\approx3.2\xi_c$ ($\sim2.8\xi_c$ in the toy model; the shift from the numerical calculation is due to different transparencies)  resulting from the diminished presence of $m\neq 0$ triplet pair correlations. The $0-\pi$ transition is recovered in curve (E) by fixing $d_{F_3}$ (see text). Parameters are $(\phi_{\mathrm{F}_1},\phi_{\mathrm{F}_2},\phi_{\mathrm{F}_3})=(0,\pi/2,0)$, $h=(3,14,3)\pi T_c$, $T=0.4T_c$ with $d_{F_2}=6\xi_c$.
}
\end{figure}
Fig.~\ref{3Fcurrentsfocus} presents the Josephson current through the perpendicular trilayer spin valve ($\phi_{F_1} = \phi_{F_3} = 0$, $\phi_{F_2} = \pi/2$), using both the full numerical calculation and the toy model. To calculate the current in the toy model, and check that it reproduces the qualitative features of the full numerical determination of $I_c$, we use Eq.~\eqref{jcdensity} replacing $f_\alpha\rightarrow\mathcal{F}$ from Eq.~\eqref{LRTCgorkovfct}.\cite{bakerPRB16,bakerEPL14} Line A in Fig.~\ref{3Fcurrentsfocus} (dashed black line; same as the $\phi=\pi/2$ line in Fig.~\ref{S3FS_current}) is obtained from the full numerical computation and is compared with line C for the toy model (solid blue line). They show that the toy model correctly captures the experimentally measurable Josephson current of the 3F spin valve system, giving similar magnitude and DTM of the current albeit shifted to smaller values of $d_{F_1}$: the DTM appears at $d_{F_1,F_3}/\xi_c \sim 2.8$ in the toy model (lines C,D) and $d_{F_1,F_3}/\xi_c \sim 3.2$ in the full numerical calculation (lines A, B).

As discussed in the previous section an examination of the Gor'kov function in Fig.~\ref{toymodelgorkov}b demonstrates why the dip appears in Fig.~\ref{3Fcurrentsfocus} and why it is not a $0-\pi$ transition. The absence of asymmetry for identical F$_1$ and F$_3$ implies that the contributions to the Josephson current of correlations from the left and right S are constructive and do not lead to a node and change of sign of the current as a function of $d_{F_1}(\equiv d_{F_3})$. For $d_{F_1}$ corresponding to the bottom of the current dip, the $m=0$ correlations have a node located exactly at the F$_1$F$_2$ (F$_2$F$_3$) interface for pairs leaking from the left (right) superconductor (see Fig.~\ref{toymodelgorkov}). This results in the lowest amount of $m\neq 0$ correlations in F$_2$ contributing to $I_c$.

The different locations of the DTM in Fig.~\ref{3Fcurrentsfocus} (curves A,C and B,D) result from the simplification made in the toy model where only correlations essential for the physics described are taken into account, and different transparencies are considered: slightly opaque in the toy model and transparent F interfaces in the numerical calculation.

The decrease of the current but absence of its reversal is the characteristic feature of the DTM, and is a distinctive feature of the presence and the role that $m\neq 0$ components play in the generation of the Josephson current through the spin valve structure. In order to tune the location of the DTM, an experimental setup may vary the magnetization strengths, $h$, of the F\1 and F\3 layers.

\paragraph*{From the DTM to the $0-\pi$ transition.} It is natural to ask how  the DTM can be transformed into a $0-\pi$ transition. The only requirement for this to happen is to induce an asymmetry between F$_1$ and F$_3$. It is shown in Ref.~\onlinecite{houzetPRB07} that a calculation with $d_{F_1}=d_{F_3}$ but changing the magnetization direction of either F\1 or F\3 (the relative canting angle) reverses the current. If this operation is performed, then the correlations generated from the left and right S acquire opposite sign and the sign of the currents in Fig.~\ref{3Fcurrentsfocus} change at the minimum; the minimum transforms into a node. The second way to transform the DTM into a $0-\pi$ transition is debuted here in Fig.~\ref{3Fcurrentsfocus}, curve E. This occurs by varying, say $d_{F_1}$, while keeping $d_{F_3}$ constant. When $d_{F_1}$ crosses a node of the $m=0$ triplet component passing the dip near $x/\xi_c \sim 3$, then the $m\neq0$ components from the left side change sign while those from the right remain unchanged. This case is depicted as curve E (dash-dotted green line) in Fig.~\ref{3Fcurrentsfocus}. We emphasize that the curves A, C and E have similar shape on the figure, but only curve E displays an actual current reversal.  This $0-\pi$ transition is still of the Houzet-Buzdin type, but from a different mechanism from Ref.~\onlinecite{houzetPRB07}. The $0-\pi$ transition is similar to the transition of the 5F presented in Sec.~\ref{ss:pentalayerJc}, since the sign change of the $m\neq0$ components is controlled by $m=0$ components (See the discussion in Sec.~\ref{ss:pentalayerJc}).

We end this section by pointing out the effect on the DTM when canting the central layer F$_2$ away from $\pi/2$. The minimum of the DTM deepens in Fig.~\ref{S3FS_current} as one decreases the canting of F$_2$. Below a certain angle $\phi$ the minimum will cross the $I_c=0$ line. When that happens, two $0-\pi$ transitions appear along the $d_{F_1}/\xi_c$ axis.  The separation between these transitions increases as one decreases the canting from the $\phi = \pi/2$ case since the algebraic minimum continues moving down. As one further lowers the canting the DTM transforms into a true $0-\pi$ transition for a very small canting angle interval $0\leq \phi<\pi/8$. This feature, implied by Fig.~\ref{S3FS_current}, underlines the point made earlier about the drastic changes of pair correlations as one approaches the homogeneous configuration. Interestingly, the $0-\pi$ transitions generated in this way result from the competition between the increasing $m\neq0$ components and the $m=0$ correlations and are thus of the singlet-triplet type according to the classification proposed in Ref.~\onlinecite{bakerPRB16}. Noteworthy is that this transition occurs in a discrete domain wall, adding to the classification of $0-\pi$ transitions of that paper.  It is also possible to find this behavior without canting and for highly tuned parameters choices for the trilayer.

\section{Conclusion}\label{s:conclusion}

Spin valves with three and five homogeneous but misaligned ferromagnetic layers were studied to show the effect of canting on pair correlations and the Josephson current.  While the trilayer has parallel-spin pair correlation components ($m = \pm 1$) that are robust to canting, the pentalayer is much more susceptible to misalignement. The pentalayer also tunes in zero-spin-projection pair correlations ($m=0$) components in the central layer.
Another difference between the trilayer and the pentalayer is that the presence of $m=0$ components in the central layer controls the sign of the $m\neq 0$ components generated from either side of the multilayer. As a result, the critical current in the pentalayer displays a $0-\pi$ transition (current reversal) that is determined by the $m=0$ components.

Both geometries display characteristic features related to $m\neq 0$ pair correlations for a wide range of canting angles; a physical system does not need to have perfectly perpendicular magnetization configuration for these effects to be prominent. With a canting angle that is too small, of course, the results will mimic a homogeneous system, but above the minimum angle used here of $\pi/8$, a measurable current was for example obtained in the trilayer that does not vary drastically from the $\pi/2$ configuration with canting (by a factor of less than 10). This outcome is relevant for experiments and applications.

The determination of the critical current as a function of the central layer thickness $d_{F_3}$ for varying canting in the pentalayer led to distinguish three thickness regimes. For thin central layers increased canting leads to a current approaching that of the trilayer spin valve. In the intermediate regime an increase in canting leads to the appearance of a $0-\pi$ transition, characteristic of the presence of $m=0$ pair correlations in the central layer. Finally, for large thicknesses we observed a {\it decrease} of the current with increased canting. The critical current through the pentalayer thus provides a variety of predictions to be tested experimentally.

An important result of this work is the demonstration that for wide enough junctions, depleted-triplet minima (DTM) rather than $0-\pi$ transitions appear in the critical current profile. Unlike the transition, the minimum cannot simply be identified by inspecting the figures depicting the Josephson current as a function of a parameter of the system (thickness or canting of the magnetic layer for example); it is difficult to distinguish on a logarithmic scale a dip representing a minimum from the dip characteristic of a $0-\pi$ transition node. Introducing an analytic toy model, we  showed that for wide junctions the dips indeed are true minima that do not cross the $I_c=0$ line. These minima appear when two effects occur simultaneously. A node of the $m=0$ pair correlations must be located at the interface between ferromagnets and a hidden sign change must occur for both the left and right correlations contributions (unseen in the Josephson current). The concurrence of these two effects lead to a reduction of the $m\neq 0$ triplet contributions that determine the magnitude of the Josephson current in these spin valves. It is recommended that experimental studies on spin valves with thicknesses beyond those investigated so far be made to identify the DTM feature. This experiment would provide yet another unambiguous demonstration of the appearance of $m\neq0$ components. The results obtained by varying the canting angle provide a variety of situations that could be used in applications.

%%%%%%%%%%%%%%%%%%%%%%%%%%%%%%%%%%%%%%%%%%%
\section{Acknowledgements}
We gratefully acknowledge funding provided by the National Science Foundation (DMR-1309341). T.E.B. thanks the Pat Beckman Memorial Scholarship from the Orange County Chapter of the Achievement Rewards for College Scientists Foundation.  T.E.B.~graciously thanks the postdoctoral fellowship from Institut quantique. This research was undertaken thanks in part to funding from the Canada First Research Excellence Fund (CFREF). 

%%%%%%%%%%%%%%%%%%%%%%%%%%%%%%%%%%%%%%%%%%%
\bibliography{spinvalve}

\end{document}